\documentstyle[12pt,rotate,cite,epsfig]{article}
\newcommand{\beq}{\begin{equation}}
\newcommand{\eeq}{\end{equation}}
\newcommand{\bea}{\begin{eqnarray}}
\newcommand{\eea}{\end{eqnarray}}
\newcommand{\bd}{\begin{displaymath}}
\newcommand{\ed}{\end{displaymath}}
\newcommand{\f}{\frac}

\newcommand{\be}{$B^{\pm} \rightarrow \eta^{\prime} K^{\pm}$}
\newcommand{\bsg}{$b \rightarrow s g $ }
\newcommand{\et}{$\eta^{\prime}$ }
\newcommand{\ete}{\eta^{\prime} }
\newcommand{\eb}{\epsilon_{\rm B}}

\begin{document}
\bibliographystyle{physics}
\renewcommand{\thefootnote}{\fnsymbol{footnote}}

\author{
Dongsheng Du${}^{1,2}$~~~C.S.Kim${}^3$ 
\footnote{kim@cskim.yonsei.ac.kr,~cskim@kekvax.kek.jp,~
http://phya.yonsei.ac.kr/\~{}cskim/}~~~and
~~~  Yadong Yang${}^{1,2}$
\footnote{Email: duds@bepc3.ihep.ac.cn, yangyd@hptc5.ihep.ac.cn}\\
{\small\sl ${}^{1}$ CCAST (World Laboratory), P.O.Box 8730, Beijing 
100080, China}\\
{\small\sl ${}^{2}$ Institute of High Energy Physics, Academia Sinica,
P.O.Box 918(4), Beijing 100039, China\thanks{Mailing address} }\\
{\small\sl ${}^{3}$ Department of Physics, Yonsei University, Seoul 
120-749, Korea}\\
}
\date{}
\title{
{\large\sf
\rightline{BIHEP-Th/97-15}
\rightline{SNUTP 97-150}
\rightline{YUMS 97-029}
}
\vspace{1cm}
{\LARGE\sf A New Mechanism for $B^{\pm}\rightarrow \eta^{\prime}K^{\pm}$ in 
Perturbative QCD   }}

\maketitle
\thispagestyle{empty}
\begin{abstract}
\noindent
We present a new mechanism for $B^{\pm}\rightarrow  \eta^{\prime} K^{\pm}$,
and analyze it strictly within perturbative QCD.
We find that such a new mechanism {\it may} dominate
non-leptonic $B$-decays to light mesons.
Within a reasonable parameter space, our prediction is in good agreement
with the recent CLEO data on $B^{\pm}\rightarrow \eta^{\prime} K^{\pm}$.
We conclude that there is no room for an abnormally large \bsg vertex 
from  physics beyond the Standard Model.
\end{abstract}

\newpage
\setcounter{page}{1}

\setcounter{footnote}{0}
\renewcommand{\thefootnote}{\arabic{footnote}}
Recently, the CLEO collaboration  has reported a large 
\et yield in charmless  $B$ decays as follows \cite{cleo1},
\bea
{\cal BR}( B^{\pm}\rightarrow \eta^{\prime } X_{s} )
=(6.2\pm 1.6 \pm 1.3)\times 10^{-4} ~~~~(2.0<P_{\eta^{\prime}} <2.7 GeV)~,
\eea
and a corresponding large exclusive branching fraction\cite{cleo2}
\bea
{\cal BR}(B^{\pm} \rightarrow \eta^{\prime} K^{\pm})
=(7.1^{+2.5}_{-2.1}\pm 0.9)\times 10^{-5}~.
\eea
It is another great experimental achievement in rare $B$ decays  
since the measurements
of $B\rightarrow K^* \gamma$ and $B\rightarrow X_s \gamma$ which involve the
so-called QCD- and electroweak-penguins. Since then, theoretical
investigations on these decays have appeared, offering several
interesting interpretations of the data, 
both within and beyond the Standard Model (SM) \cite{hal}.
In particular, the decay mode (2) makes  itself conspicuous due to its
{\it surprisingly  large} branching ratio. 

Now let us examine the theoretical status of the estimate of the 
exclusive branching fraction
${\cal B}(B^{\pm} \rightarrow \eta^{\prime} K^{\pm})$. 
The standard theoretical framework to study non-leptonic 
$B$ decays is based on the effective Hamiltonian which describes the 
decays at quark level,
\bea 
H_{eff}=\frac{G_F}{\sqrt 2 }\sum V_{_{CKM}} C_i {\cal O}_i ~,
\eea
and  the BSW model \cite{bsw} to estimate the hadronic matrix element
$<M_1 M_2 \mid {\cal O}_i \mid B>$.

An important feature of the BSW model in nonleptonic decays is the
use of the factorization and spectator ansatz. It works remarkably
well in the  so called heavy-to-heavy transitions because the
ansatz  is consistent with the HQET, and therefore is almost certainly true. 
However, we think that it has never been tested, and so is much less justified,
in  heavy-to-light decays.

Even given the validity of the BSW ansatz, yet another
problem exits  in estimating the hadronic matrix elements like
\bea
<\ete \mid \bar u (1-\gamma_5 ) b\mid B^- > ,~~
&<K^- \mid \bar s (1+\gamma_5 )u \mid 0>, \nonumber\\
<\ete \mid \bar s \gamma_5 s \mid 0>,~~~  
&<\ete \mid \bar u\gamma_5 u \mid 0>.
\eea
The Dirac equations of motion
\bea
\bar q q^{\prime} =\frac{-i\partial^{\mu}(\bar q \gamma_{\mu}q^{\prime})}
{m_q -m_{q^{\prime}}}~,~~~~~~~ 
\bar q\gamma_5 q^{\prime}=\frac{-i\partial^{\mu}(\bar q \gamma_{\mu}
\gamma_5 q^{\prime})}
{m_q +m_{q^{\prime}}}
\eea
are used to get the elements for
$B^{\pm}\rightarrow \ete K^{\pm}$,
\bea
 <\ete \mid \bar s \gamma_5 s \mid 0>&=i\frac{ m^2_{\ete}}{2m_s}f^{(s\bar s)}\\
 <\ete \mid \bar u \gamma_5 u \mid 0>&=i\frac{ m^2_{\ete}}{2m_u}f^{(u\bar u)}\\
 <K^- \mid \bar s (1+\gamma_5 )u \mid 0>&=i\frac{ m^2_{\ete}}{m_s +m_u} f_K~.
\eea
With these relations, the amplitude for  $B^{\pm}\rightarrow \ete K^{\pm}$ 
reads
\bea
{\cal M} =
\frac{G_F}{\sqrt{2}} 
\{
V_{ub}V^{\star}_{us}
\left[ 
a_2 +a_{1}\frac{ M_B^2 -m_{\ete}^2 }{ M_B^2 -m_K^2 }
\frac{F_0^{B\rightarrow \ete}(m_K^2 )}{F_0^{B\rightarrow K^-}(m_{\ete}^2)}
\frac{f_K}{f_{\ete}}
\right]
\nonumber\\
-V_{tb}V^{\star}_{ts}
\left[ 
2a_3 -2a_5 +( a_3 -a_5 +a_4 +\frac{a_6 m_{\ete}^2}{m_s (m_b -m_s )} )
\frac{f^{(s\bar s)}_{\ete}}{f^{(u\bar u)}_{\ete}}
\right. \nonumber\\
 \left.
+(a_4 +\frac{2a_6 m_K^2 }{(m_s +m_u )(m_b -m_u )})
\frac{M_B^2 -m^2_{\ete}}{M_B^2 -m^2_K }
\frac{F_0^{B\rightarrow \ete}(m_K^2 )}{F_0^{B\rightarrow K^- }(m_{\ete}^2)}
\frac{f_K }{f^{(u\bar u)}_{\ete}}
\right]
\}\nonumber\\
<K^{-}\mid \bar{s} \gamma_{\mu}(1-\gamma_5 )b\mid B^- >
<\ete \mid \bar{u}\gamma_{\mu}(1-\gamma_5 )u\mid 0>~.
\eea
Assuming the Dirac equation is valid for bounded fermions, the
factor $\frac{m_{\ete}}{m_s} $ in the amplitude enhances the contribution
of operator ${\cal O}_6$, which is very difficult to understand.    

In an alternative way, we turn to estimate the amplitude using the
perturbative QCD hard scattering formalism \cite{lb}, which has also been
extensively applied to $B$-decays \cite{lcb}. The diagrams to be
calculated are depicted in Fig. 1. However, we find that these
amplitudes are rather small due to cancellations between the diagrams
and the small $b\rightarrow g g $ vertex \cite{simma}. Nevertheless,
before one goes beyond the SM, the contributions in the SM should be
 examined carefully and exhausted.  In what follows, we discuss a new
type of mechanism for $B$-decays to light mesons in detail.

The new mechanism is depicted in Fig. 2. This mechanism is motivated by 
the fact that both the recoil between \et and $K^-$ and the
energy released in the process
are  large. The gluon from \bsg vertex  carries energy   about $M_B /2$
and then materializes to $\eta^{\prime}$ and emits another {\it hard}  gluon 
to balance color and momentum.
The    momentum squares of the gluons scale as~ ${\propto}M_B^2$
\bea
k_1^2 &=((1-x)P_B -yP_K )^2 \cong M_B^2 -(1-x)y(M_B^2 -M_{\eta^{\prime}}^2 )~,
\nonumber\\
{\rm and}k_2^2 &=(x P_B -(1-y)P_K )^2 \cong -x(1-y) M_B^2 ~,
\eea
where $x$ and $y$ are the momentum fractions carried by
the  collinear quarks shown in Fig. 2. For self-consistency of the co-linear 
picture used here, the terms $\sim x^2 M_B^2$ 
are neglected since  they are at the same level of the transverse momentum
square of the quarks in the bound state $B$ meson. 
Using mean values $<y>\sim \frac{1}{2}$, $<x>\sim \epsilon_{_B}$ with
$\epsilon_{_B} \sim 0.05-0.1$ \cite{lcb}, we get $<k_1^2 >\sim 12$ GeV$^2$ and
$<k_2^2 >\sim 1$ GeV$^2$, which are large enough to justify
a perturbative calculation.     

The soft contributions are parameterized in terms of wave functions of the
bound states. In the spirit of Ref. \cite{lb}, we neglect the transverse 
components of quarks  and take  the wave functions for $B^-$ and $K^-$ as 
\bea 
\Psi_B &=\frac{1}{\sqrt{2}} \f{I_C }{\sqrt{3}} \phi_{B}(x)
(\slash{\hskip -3mm}P_B  
+M_B )\gamma_{5} ~, \nonumber\\
\Psi_{K} &=\frac{1}{\sqrt{2}} \f{I_C }{\sqrt{3}} \phi_{K}(y)
\gamma_5 ~~\slash{\hskip -3mm} q_{K} ~,
\eea
where $I_C$ is an identity in color space. In QCD, the integration of the
distribution amplitude is related to the meson decay constant 
\bea
\int \phi_K (y)dy=\frac{1}{2{\sqrt 3}}f_K ~,~~~~
\int \phi_B (x)dx=\frac{1}{2{\sqrt 3}}f_B ~.
\eea
We write down the amplitude of Fig. 2 as 
\bea
\label{m}
{\cal M}=\int dxdy 
\phi_B(x) \frac{Tr \left[\gamma_5~\slash{\hskip -3mm} q~~ \Gamma_{\mu}
(~\slash{\hskip -3mm}p  +M_B )\gamma_{5}\gamma_{\nu} \right]
4\epsilon^{\mu\nu\alpha\beta}k_{1\alpha}k_{2\beta} C_{eff}C_{F} g_s^3}
{{\sqrt 2}{\sqrt 2} (k_{1}\cdot k_2) k_1^2 k_2^2 }\phi_K(y) ~,
\eea
where $\Gamma_{\mu}$ is the effective \bsg vertex known for years \cite{hou88}
\bea
\Gamma^a_{\mu}=\frac{G_F}{\sqrt{2}}\frac{g_s}{4\pi^2}V^{\star}_{is}V_{ib}
t^a \{ F_1^i (k_1^2 \gamma_{\mu}-k_{1\mu}~~\slash{\hskip -3mm}k_1 )L
-F_2^i i\sigma_{\mu\nu}k_1^{\nu} m_b R \} ~.
\eea
We have used a Lorentz and gauge invariant amplitude
$<g^a (k_1 ,\epsilon_1 )g^b (k_2 ,\epsilon_2 ) \mid \ete>$ in Eq. (\ref{m}), 
with the shorthand notation
$G_{\mu\nu}\sim k_{\mu}\epsilon_{\nu}-\epsilon_{\mu}k_{\nu}$, and it reads
\cite{kuhn, close}
\bea
<g^a g^b \mid \ete>=\delta^{ab}A_{\ete}F(k_1^2 ,k_2^2 )
\epsilon^{\mu\nu\rho\sigma}G^1_{\mu\nu}G^2_{\rho\sigma}
\eea
with $A_{\ete}=\frac{2C_{eff}}{m^2_{\ete}}$
and the form factor $F(k_1^2 ,k_2^2 )=\frac{m^2_{\ete}}{2k_1 \cdot k_2 }$
which will be extracted form the $J/\Psi$ decays.
$C_F$ is a color factor 
$C_F =\frac{1}{\sqrt{3}}\frac{1}{\sqrt{3}}Tr[T_a T_b ]\delta_{ab}=\frac{4}{3}$.     

Finally we obtain 
\bea
{\cal M}=\frac{G_F}{\sqrt{2}}\alpha_s^2 C_{eff}C_F 32 (V_{is}V_{ib}^{\star})
~~~~~~~~~~~~~~~~~~~~~~~~~~~~~~~~~~~~~~~~~~~~~~~~~~~~~~~~~~~~~~~~\nonumber\\
\times\int dxdy\phi_{B}(x)\phi_{K}(y)
\frac{
F_1^i k_1^2 [p\cdot k_1 q\cdot k_2 -p\cdot k_2 q\cdot k_1 ]
+F_2^i M_B m_b [q\cdot k_2 k_1^2 -q\cdot k_1 k_1 \cdot k_2 ] }
{k_{1}\cdot k_2  k_1^2 k_2^2 }, 
\eea
where the momenta $k_1$, $k_2$, $p$, $q$ can be read off from Fig. 2.
We note  that from Eqs. (10,16) the singularities of 
the gluons-propagators are located at the point $y=1$, which  
is just the end point for the wave function $\phi_{K}(y)$.
However, it is well known that the value of bound state 
wave function at its end point
is exactly zero in QCD.

We can extract  $C_{eff}$ form the data  $J/\Psi \rightarrow \ete \gamma $
to remove ambiguities:
\bea
\frac{ \Gamma(J/\Psi \rightarrow \ete\gamma)}
{\Gamma(J/\Psi \rightarrow e^+ e^- )} 
=\frac{4}{\pi}\frac{\alpha^4_s (m_{\Psi})}{\alpha_e}\frac{1}{M^2_{\Psi}}
(8\sqrt{3}C_{eff})^2 \frac{x \cdot \mid{\hat H}^{ps}(x)\mid^2}{54} ~,
\eea
where the $x \cdot \mid{\hat H}^{PS}(x)\mid^2 $ can be found in \cite{kuhn}. 
For 
$x=1-\frac{m^2_{\ete}}{M_{\Psi}^2}$, 
$x \cdot \mid {\hat H}^{PS}(x)\mid^2 =54.8$. We extract $C_{eff}=0.075$ GeV.
In order to get quantitative estimates, we take the wave function 
as \cite{lcb}
\bea
\phi_B (x)=\frac{f_B}{2\sqrt{3}}\delta(x-\epsilon_B )~,~~~~~{\rm and}~~~~~
\phi_K (y)=\sqrt{3} f_{K} y(1-y)
~,
\eea
and $\tau_B =1.62~ps$, $f_B =140$ MeV, $f_K =113$ MeV (corresponding
$f_{\pi}=92$ MeV) and
$\mid V_{ts}V_{tb}\mid=0.044$. For the strong coupling at the energy scale
$<k_2^{2}>$,
we respect the choice $\alpha_s=0.38$ in Ref. \cite{lcb}.
The main uncertainty of our estimate exists in the
distribution function $\phi_B (x)$. At present, it can not be derived
from the first principle of QCD. 
However, it is easy to understand that the distribution
function is peaked sharply at one  point due to the heavy $b$ quark
carrying most of the momentum of the $B$ meson. 
The peaking position is expected at
$x=\frac{M_B - M_b }{M_B } \sim \frac{\Lambda_{QCD}}{M_B }$.

From ${\it Eq}.10, 16, 18$, it is easy to see that the integration in
${\it Eq}.16$ is infrared safe because the pole in $1/k_2^2 $ is canceled
by the $K$ meson wavefunction and $k_1^2 $ is always positive. To confirm
that $\mid {\cal M}\mid$ is dominated by hard gluon contributions,
we  define
$R$ as the ratio of $\mid {\cal M}\mid$ with cut on $k_2^2$ to that
without cut. Taking $\Lambda_{QCD} \cong 200MeV$, we list the numerical
results of $R$ in Table I to show that the integration in ${\it Eq}.16$ is  
indeed  dominated by hard gluon contributions.
For somewhat complete investigation on this mechanism, 
the next to leading order contributions(high twist) should be included,
however, it is complicated. We shall discuss it elsewhere.    

The  numerical results of  ${\cal BR}( B^{\pm} \rightarrow \ete K^{\pm})$
are displayed in Fig. 3 as a function of the peaking position parameter
$\epsilon_{_B} $.

We can see that
our predictions are in a good agreement with experimental results
in the region of $\epsilon_{_B} =0.05 \sim 0.07$.
Furthermore, if the contributions  of Eq. (9) estimated from the conventional
way (which may contribute up to
${\cal BR}( B^{\pm} \rightarrow \ete K^{\pm}) = (1\sim 4)
\times 10^{-5} $ 
\cite{ali,cheng97}) are taken into account, the SM predictions turn
out to be in agreement with the CLEO data within the $1\sigma$ level
in the whole parameter range of $\epsilon_{_B} = 0.05 \sim 0.1$.  
We conclude that
${\cal BR}( B^{\pm} \rightarrow \ete K^{\pm})$ is {\em not} ``surprisingly 
large'', and the mechanism in the Standard Model 
presented here seems sufficient for explaining it.     

\vspace{0.5cm}
{\it Note added:} After finishing this paper, we became aware that a similar 
idea appeared in hep-ph/9710509 \cite{arhmady}. But our physical picture, 
calculation method and conclusion are all different from theirs, which 
used a gluon-diffusion picture and calculated it by using  effective
Hamiltonian 
method. However, we here present  a hard scattering picture and solve it
on perturbative QCD.
\vspace{0.5cm}
\noindent
{\large\bf Acknowledgment}
\noindent
We thank H.Y. Cheng and M. Drees for careful reading of the manuscript and 
valuable comments.  D.Du and Y.D.Yang are supported 
in part by the National Natural Science Foundation and the 
Grant of State Commission of Science and Technology of China.
The work of CSK was supported 
in part by Non-Directed-Research-Fund, KRF in 1997,
in part by the CTP, Seoul National University, 
in part by Yonsei University Faculty Research Fund of 1997, 
in part by the BSRI Program, Ministry of Education, Project No. BSRI-97-2425
and in part by the KOSEF-DFG large collaboration project, 
Project No. 96-0702-01-01-2.

\newpage

\begin{figure}[tb]
\vspace*{-5cm}
\hspace*{-2cm}
\centerline{\epsfig{figure=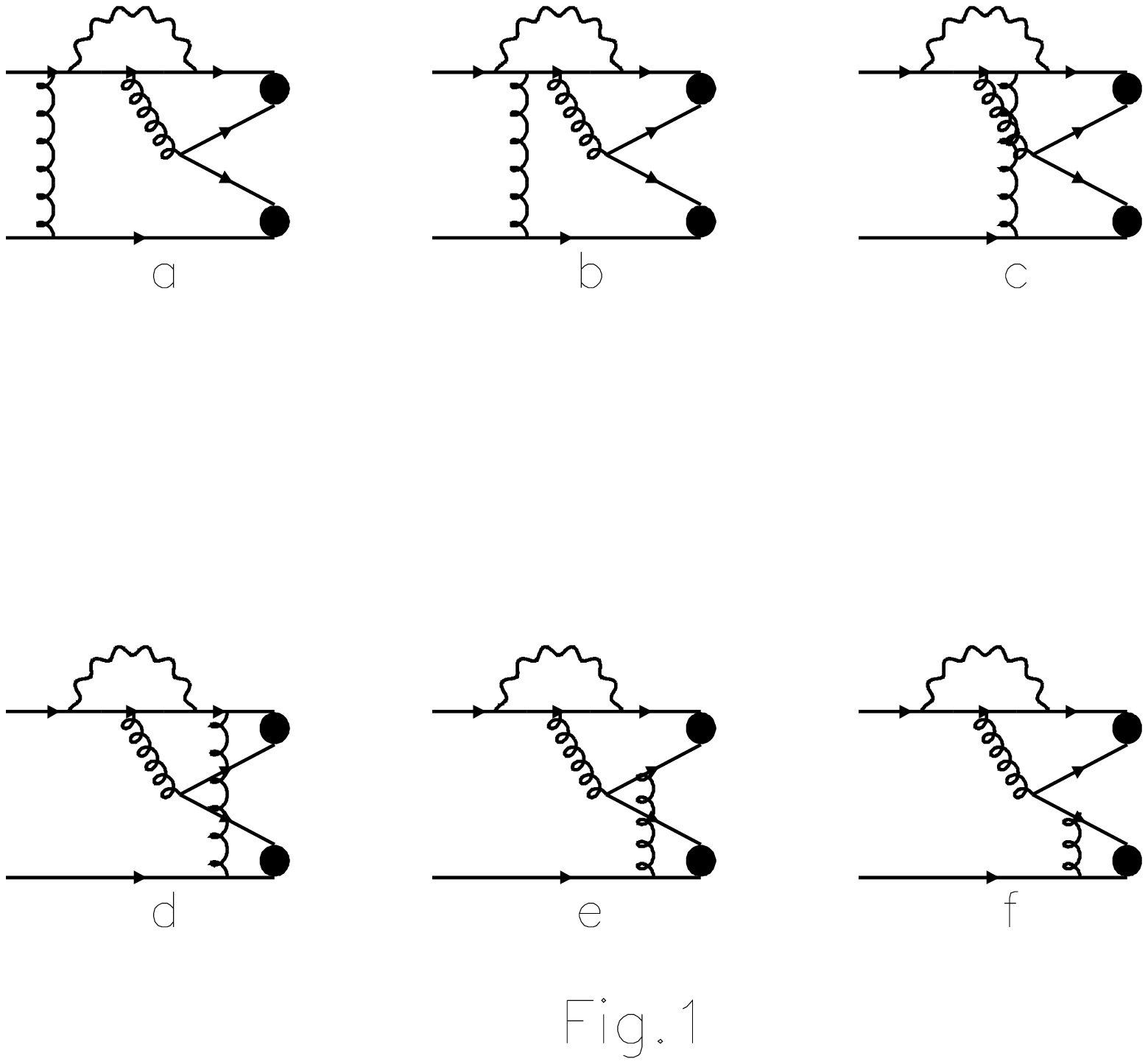,height=18cm,width=17cm,angle=0}}
\vspace*{-2cm}
\caption{\em Time and space like strong penguin diagrams for \be in usual way.
Diagrams suppressed by $V_{ub}$ are neglected. Blobs depict  $K$ and
$\eta^{\prime}$ }
\end{figure}
\begin{figure}[tb]
\vspace*{-5cm}
\hspace*{-2cm}
\centerline{\epsfig{figure=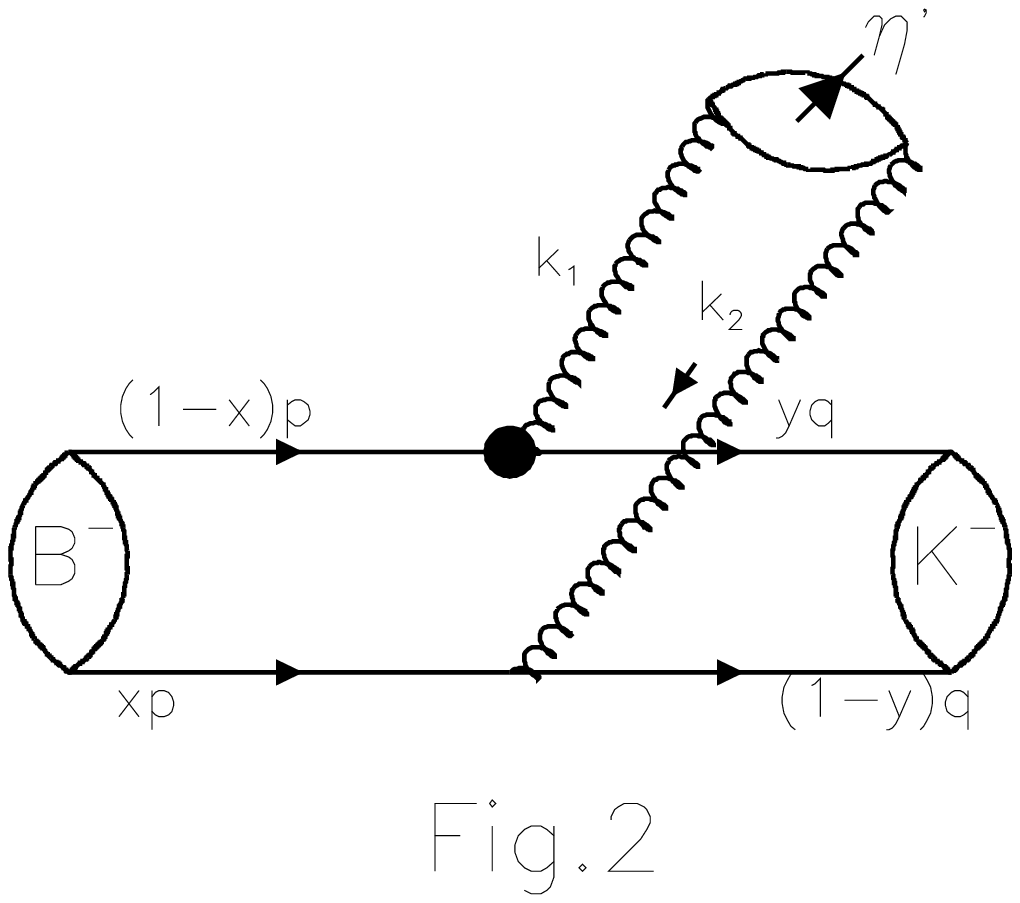,height=18cm,width=17cm,angle=0}}
\vspace*{-3cm}
\caption{\em  New diagram for \be. $x,~ 1-x,~ y$  and $1-y$ are momentum fractions 
carried by the quarks.}
\end{figure}
\begin{figure}[tb]
\vspace*{-5cm}
\hspace*{-2cm}
\centerline{\epsfig{figure=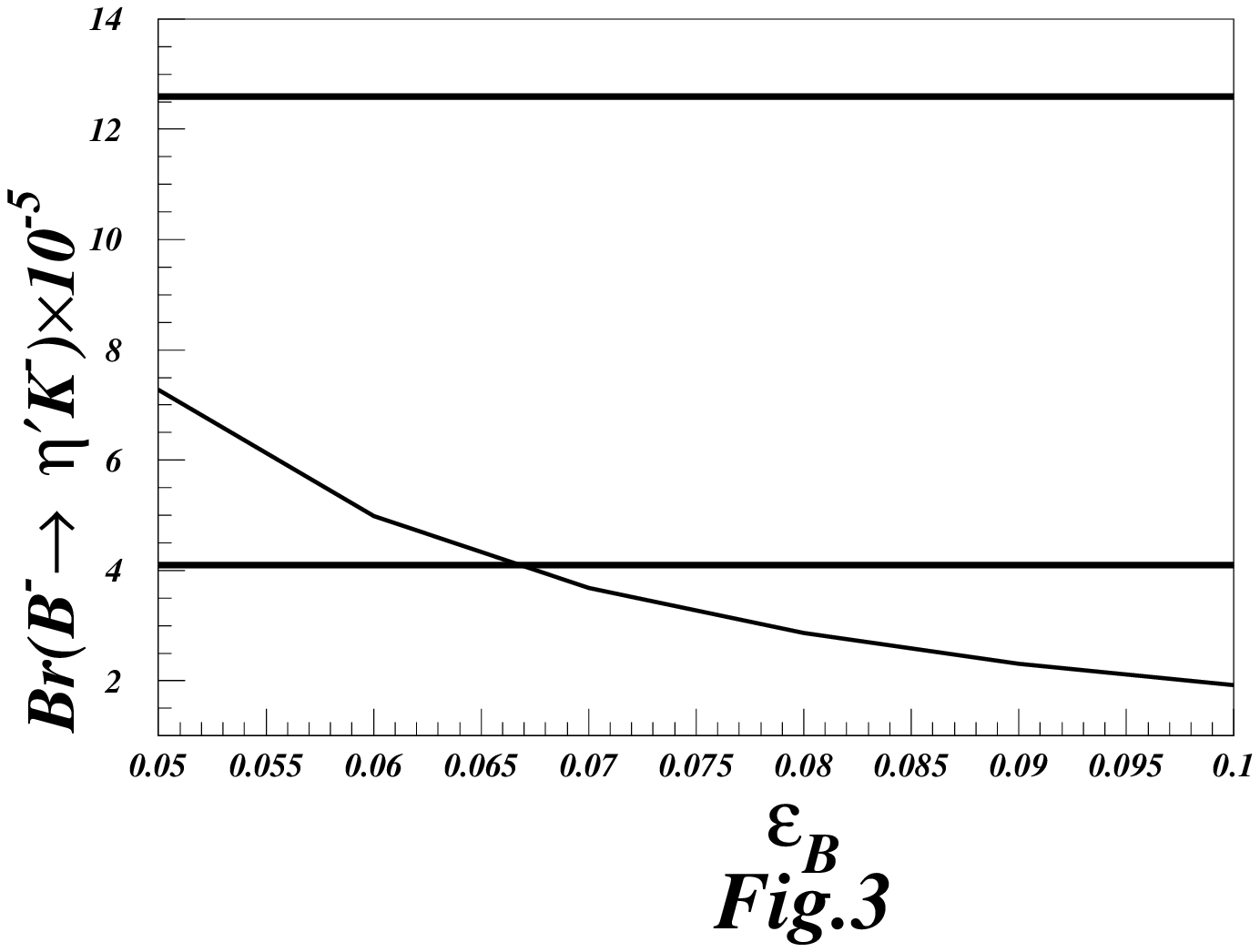,height=18cm,width=17cm,angle=0}}
\vspace*{-3cm}
\caption{\em  ${\cal BR}$(\be) as a fuction of $\epsilon_{_B}$. 
The horizontal thick solid lines show the CLEO measurement
(with $\pm 1\sigma$ error bar) and the thin curve is the
contribution from the new mechanism presented here.}
\end{figure}
\begin{table}[tb]
\vspace*{9cm}
\hspace*{-1cm}
\centerline{\epsfig{figure=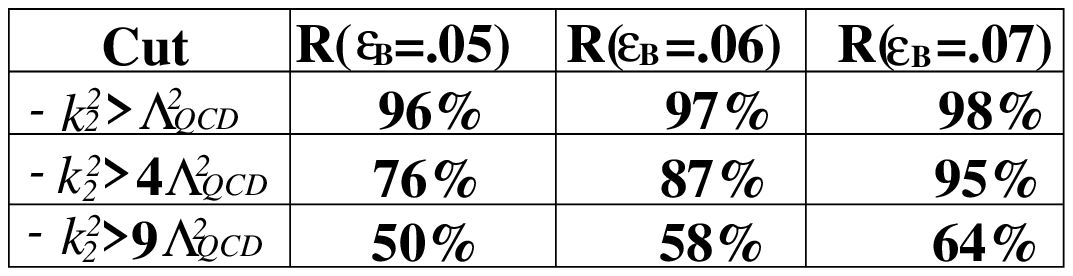,height=14cm,width=4cm,angle=-90}}
\vspace*{2cm}
\caption{\em  Numerical results for $R$ for different cuts and $\epsilon_B$ .}
\end{table}
\end{document}